\documentclass[runningheads]{llncs}
\usepackage{graphicx}
\usepackage{hyperref}
\usepackage{cite}
\usepackage{booktabs}
\usepackage{float}
\usepackage{enumerate}
\usepackage{verbatim}
\usepackage{enumitem}
\usepackage{xcolor}
\usepackage{framed}
\usepackage[T1]{fontenc}

\begin{document}
\authorrunning{Duc-Tien Dang-Nguyen et al.}
\titlerunning{How Common Media Platforms Handle Uploaded Images}

\title{Practical Analyses of How Common Social Media Platforms and Photo Storage Services Handle Uploaded Images}

\author{
    Duc-Tien Dang-Nguyen \inst{1,2}
    \and Vegard Velle Sjøen\inst{1}
    \and Dinh-Hai Le\inst{3,4} 
    \and \\ Thien-Phu Dao\inst{3,4}
    \and Anh-Duy Tran\inst{3,4}
    \and Minh-Triet Tran\inst{3,4}
}
\institute{
    University of Bergen, Norway
    \and Kristiania University College, Norway
    \and University of Science, Ho Chi Minh City, Vietnam
    \and Vietnam National University, Ho Chi Minh City, Vietnam
}
\maketitle

\begin{abstract}
The research done in this study has delved deeply into the changes made to digital images that are uploaded to three of the major social media platforms and image storage services in today's society: Facebook, Flickr, and Google Photos. In addition to providing up-to-date data on an ever-changing landscape of different social media networks' digital fingerprints, a deep analysis of the social networks' filename conventions has resulted in two new approaches in (i) estimating the true upload date of Flickr photos, regardless of whether the dates have been changed by the user or not, and regardless of whether the image is available to the public or has been deleted from the platform; (ii) revealing the photo ID of a photo uploaded to Facebook based solely on the file name of the photo.

\end{abstract}

\section{Introduction}
Over the years, various tools and techniques for digital image forensics have been developed in the multimedia research community for things such as image manipulation detection\cite{ tran2022dedigi}, deep-fake detection \cite{mehta2021fakebuster, vo2022adversarial}, and source identification \cite{ sandoval2014source}, which all try to answer two main questions such as "Was the image captured by the device it is claimed to be acquired with?" and "Is the image still depicting its original content?" \cite{redi2011digital}. These tools and techniques, however, only work well in controlled settings such as laboratory experiments and often fail to provide reliable answers in the wild. One key factor that introduces these tools' poor performance is the processing of asocial media platforms or online storage on uploaded images \cite{ pasquini2021media}. This process might change the filename or the quality as well as the properties of an image by compressing. Thus, performing forensics on an image that is retrieved from a social network can be affected by the mentioned processes.   Therefore, knowing how the uploaded images were processed will play a key role in improving the performance and accuracy of digital forensics tools. 

In this study, we aim to investigate how social networks alter cross-posted images in the uploading process, i.e., \textbf{investigating what happens to an image in the process of being uploaded to a common platform or photo storage service, and what kind of alterations are made to the image file}. Once the modifications are understood, we will attempt to predict which digital forensics and verification tools will perform best on those images in backtracking their origin, as well as detecting tampering and manipulation. 
%


Studies have shown that social media platforms typically leave some digital fingerprints on the uploaded image itself \cite{ferreira2020review, tran2022dedigi} and the fingerprints may be in the form of changing the name of the image based on timecode, like Facebook, or adding a new portion on the bottom of the image to indicate its platform of origin, like Reddit. They can be exploited to identify the real image origin. Since every modification of an image will alter its properties, image forensics techniques’ effectiveness depends on what kind of processing an image went through. The data gathered in this paper will be used to determine how to apply different techniques for predicting the upload origin of an image. Being able to identify how an image was altered can therefore provide a significant advantage when attempting to apply different image forensics techniques.

Social media and online storage services are popular places where ordinary users can post their daily life images and upload photos. \textit{Photutorial}\cite{Photurial} data indicates that 1.2 trillion photos were shot globally in 2021. In 2022, 2.1 billion images were shared per day on Facebook and 1 million on Flickr, making Facebook and Flickr in the top list of social media for image sharing, along with WhatsApp, Instagram, and Snapchat \cite{Photurial}. Google Photos is an online app for storing and managing personal photos and videos. In 2020, Google reported that the platform had more than 4 trillion photos in its database, and 28 billion new pictures and videos are added weekly \cite{blog_google}. From the given numbers, we decided to select two social media platforms, Facebook and Flickr, and one online storage, Google Photos, for performing our research on the change in uploaded images. Inspecting favored platforms can introduce a general notion of processing online images.

The main contributions of this study are:
\begin{itemize}
\item Providing updated data on how popular media networks and photo storage services, in particular Facebook, Flickr, and Google Photos, handle user uploaded images;
\item  Proposing a novel approach to estimate the true upload date of images that have been uploaded to the Flickr platform;
\item  Proposing a novel approach for reverse engineering the image ID of an image uploaded to Facebook.
\end{itemize}

The content of this paper is organized as follows. In Section \ref{sec:relatedwork}, we briefly provide background information and related work. The methodology is discussed in Section \ref{sec:method}. The results of the research are reported in Section \ref{sec:results}. We give some discussions about the potential applications in Section~\ref{sec:applications} while Section~\ref{sec:conclusion} summarises the entire paper. 

\section{Background and Related Work}
\label{sec:relatedwork}

\subsection{Background}

\textbf{Digital Image Forensics (DIF)} works in tandem with "Digital Watermarking" to identify and contrast malicious image manipulation \cite{  tran2022dedigi}. Various approaches, from passive to active, can be applied to accomplish the DIF's task. The simplest method that could be noted is the analysis by inspecting the metadata of an image. This kind of evidence depicts an overview hint of what has been done on a particular image and then acts as a starting point for further analysis.

Apart from data used for rendering digital images, many image formats also maintain other data for describing the content or features of a file. This type of data is called metadata. We present here brief descriptions of some commonly used metadata:

\begin{itemize}[leftmargin=*, align=left, wide=0pt]
    \item \textbf{Exif Data.}
    The Exchangeable Image File Format (Exif) is a standard that specifies detailed information about a photograph or other piece of media captured by a digital camera or a mobile phone.It may also store critical information like camera exposure, date/time of image capture, and even GPS location.
    
    \item \textbf{IPTC Data.}
    IPTC metadata is essential for effective image management and preservation. It is especially valuable for news companies and journalists, image archives, and other organizations that need a complete representation of image details. Unlike EXIF data, which is concerned with the technical aspects of an image, the IPTC standard is concerned with the content of the photos, as well as their ownership, rights, and license status.
    
    \item \textbf{XMP Data.}
    Extensible Metadata Platform (XMP) is a file labeling system developed by Adobe that allows metadata insertion into files throughout the content development process, such as titles and descriptions, searchable keywords, and up-to-date author and copyright information.
    
    \item \textbf{Current IPTC Digest.}
    ExifTool by Harvey \cite{harvey2013exiftool} provides a hashing function to map data of arbitrary size to fixed-size values. This value is called Current IPTC Digest. The main usage of Current IPTC Digests is to compare to other IPTC Digests to see if a file has been modified. These hashes are generated by calculating the MD5 hash from the legacy IPTC IIM values. It is then used to compare with IPTC Digest in the metadata. However, the IPTC IIM format is no longer actively maintained since many picture management programs have moved on to IPTC XMP standards.
        

\end{itemize}

\subsection{Related Work}

This section explores relevant techniques and challenges on some digital image forensics tasks, including platform provenance, source verification, and date/time verification.
\begin{itemize}[leftmargin=*, align=left, wide=0pt]
    \item \textbf{Platform provenance:} One of the most important step in verifying user-generated content is provenance. This process aims to determine the most recent internet platform or social media network where the content was uploaded. 
    Bharati et al. \cite{bharati2019beyond} proposed a method that utilizes non-content-based information to extract the path of a particular image that has gone through the Internet without the sizeable computational overhead. Siddiqui et al. \cite{siddiqui2019social} introduced a model to find these distinct traces by utilizing Deep Learning based approaches to look at the image's social network of origin, focusing on determining which social network the particular image was downloaded from.
    \item \textbf{Source verification:} While provenance traces back to the first uploader of the image, source verification determines the context of its creation (e.g., the one who took the picture). This process also plays a key role in digital forensics since the uploader and creator may not always be the same person. Kee et al. \cite{kee2011digital} analyzed JPEG headers to identify camera signatures consisting of information about quantization tables, Huffman codes, thumbnails, and Exif. They showed this signature is highly distinct across 1.3 million images, spanning 773 cameras and cellphones. Marra et al. \cite{marra2017blind} proposed an algorithm for blind PRNU-based image clustering using noise residuals to cluster the source of an image. Mullan et al. \cite{mullan2019forensic} looked into the possibility of linking photographs from modern smartphones using JPEG header information. Their work showed that Exif metadata change is less connected to the actual Apple hardware but more closely connected to the change of version in the iPhone operating system iOS.
    \item \textbf{Date \& time verification:} In digital forensics, the time an image is taken or uploaded can be valuable information for fact-checkers \cite{ moltisanti2015image}. For example, if an image is claimed to have been taken from a specific event, while in fact the image is already uploaded before that event happens. Therefore, this claim is wrong. However, such information is proven to be significantly more difficult to extract \cite{riggs2018image}. Provenance can help indicate when is the image first uploaded. However, this does not help in determining when the material was collected. Metadata can be extracted, but it's not reliable, since it's trivial to alter these data with correct applications \cite{acker2018data}.
\end{itemize}

\section{Methods}
\label{sec:method}

\begin{figure}[t!]
    \centering
    \includegraphics[width=\linewidth]{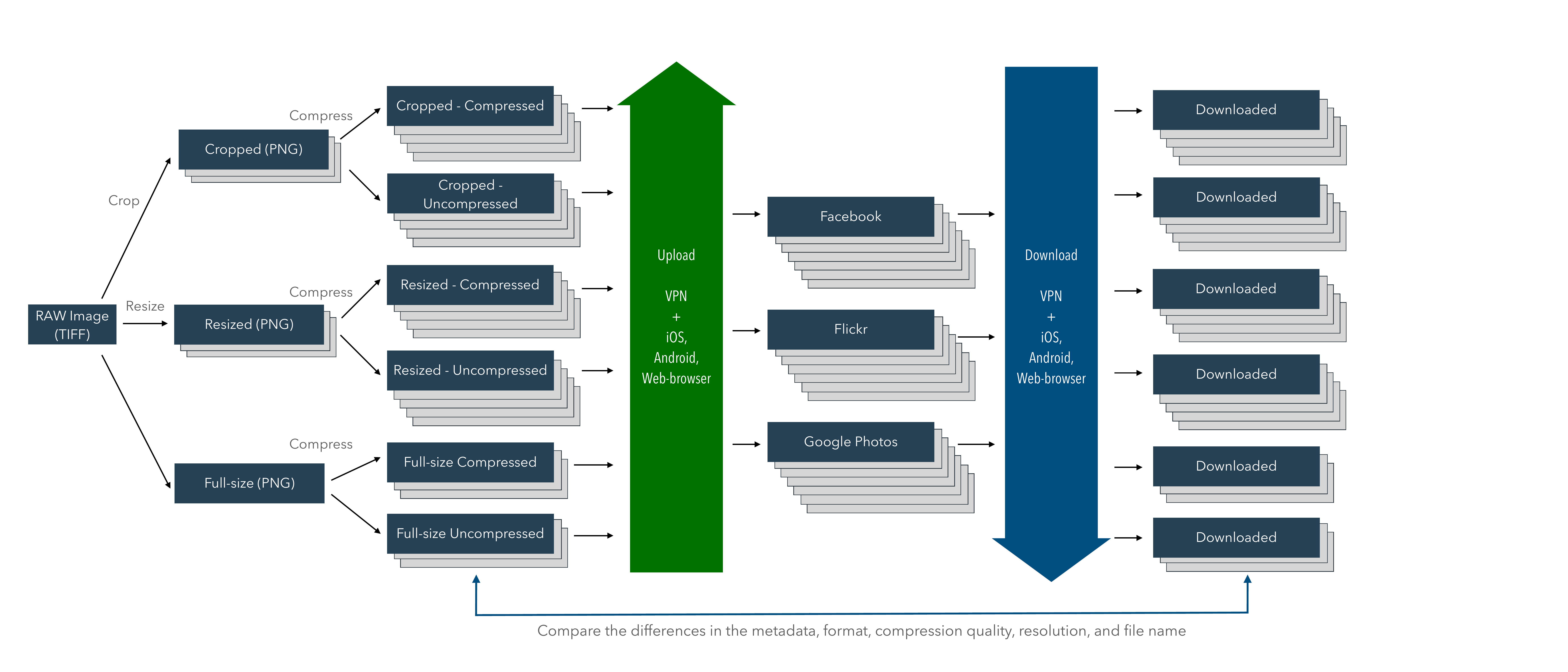}
    \vspace{-0.5cm}
    \caption{The analysis process. Starting from a raw image in TIFF format (from RAISE 1K dataset \cite{dang2015raise}), we first created different images by cropping and resizing and change the format to PNG (to be able to upload to the platforms). Each new image is then compressed under different quality factors. Finally, they are uploaded to different platforms and services via Android and iOS apps, and also via PC Web-browsers. We also tried to upload under different locations by using VPN to connect to different server in different countries. The differences before and after the uploading are then studied.}
    \label{fig:analysis}
\end{figure}

The general idea of this work is to analyse how Facebook, Flickr and Google Photos handle the uploaded images by performing various manipulations to create a plethora of specific scenarios before being uploaded to social media.  As can be seen in the Figure \ref{fig:analysis} the whole process can be split into two main phases, including before uploading to social media and downloading images from it. In the first phase, a particular image is changed by cropping, resizing, or even keeping the original. Then, this (changed) image is converted to PNG format before compressing. The final image is uploaded to all three platforms, via various ways, including native apps on iOS, Android and via Chrome browser on different PCs. We also inspect the effect of uploading from different locations by using VPN. In the second phase, we download the corresponding image through mentioned platforms and VPN from social media and then compare them with the one before uploading to inspect the changes. The detail of each manipulation process is described as follows.

\begin{itemize}[leftmargin=*, align=left, wide=0pt]
    \item \textbf{Cropping:} To analyze how online platforms react to the cropped image compared with the original one, we crop the image from 50\% - 80\% of the original size in three ways: top-left, bottom-right, and center. 
    
    \item \textbf{Resizing:} Like cropping, we also change the image size to to 200\%, 150\%, and 50\% of the original size to inspect the platform reactions. 
    
    \item \textbf{Compressing:} We also change an image's quality to examine how social media compression will apply to an uploaded image. A target image is compressed by quality factor of 80, 60, 40, and 20.
    
    \item \textbf{VPN:} To check whether social media change their behavior on uploaded images based on location or not, we leverage VPN to manipulate the root location of an image. We have tested with many countries on different continents, including Vietnam, Norway, the USA, and Australia. 
\end{itemize}    


\section{Results}
\label{sec:results}
\subsection{Data and Experimental Setup}

\textbf{Dataset:} Fifteen RAW images from the RAISE 1K dataset \cite{dang2015raise} were used. We followed the process described in Figure~\ref{fig:analysis}, and applied five different size and resolutions (ranging from $3430 \times 2278$ - to $858 \times 570$), and five different compress qualities (uncompressed, 80, 60, 40, and 20). In the end, we end up with $2475$ pairs of images before-after upload.

\textbf{Tools:}
Filenames, compression tables and resolutions were then analyzed with IBSNtools\cite{IBSNtools} and compared to the original images, as well as Exif and IPTC Data with ExifTool \cite{harvey2013exiftool}. 
We analysed filenames for any pattern that could give additional information about the image. Image resolutions were analyzed to see whether or not the images had been resized or re-compressed in the upload process. Exif and IPTC data was analyzed to see whether or not the metadata was deleted or modified.


\subsection{Results on Google Photos}

\textbf{Exif.} All Exif data was kept intact after uploading to Google Photos.

\textbf{IPTC Data.}
When uploading the JPEG images with Google Photos’ compression setting turned off, IPTC data was kept intact on all the images, resulting in an IPTC Digest hash that was identical to the original images. With the compression setting turned on, all images returned new and unique IPTC Digest hashes, meaning they still contained IPTC data, but the data had been modified for each image. 

\textbf{XMP Data}
Google Photos retained most of the XMP data when uploading images with re-compression.
Without re-compression, all the XMP data was identical to the original. In both cases, both the
"Original Document ID" and the "Preserved Filename" entries were retained and matched the
original files.

\textbf{Resolution.}
The resolution of all the images was identical to the original images, no modifications were
made to the dimensions regardless of whether or not the compression setting in Google Photos
was enabled or disabled.

\textbf{Compression.}
Unlike the other social networks, the RAW images that were uploaded were not converted to JPEG or compressed at all but remained identical to the originals. The JPEG files were recompressed, but since Google Photos offer the option to not modify the files at all, this would depend on the settings of the Google account.

\textbf{Filename.}
Filenames also remained unchanged on all the images.

\subsection{Results on Facebook}
\textbf{Exif.}
In all cases on both the JPEG and RAW originals, Facebook stripped all Exif data from the images except from the "Artist" and "Copyright" tags.

\textbf{IPTC Data.}
When uploading the RAW files, Facebook retained the same IPTC data on all images, resulting in an identical IPTC Digest on all the images. The JPEG files were completely stripped of all IPTC data.

\textbf{XMP Data.}
Facebook retained no XMP data when uploading both the JPEG and the RAW images.

\textbf{Compression.}
In all cases, Facebook compressed the images with similar compression tables with some variation. Average error rare is smaller than 2\%.

\textbf{Resolution.}
Facebook resized some of the images with larger dimensions, the threshold was observed at $2048$. Both the width and the height of the image can trigger a resize. This is consistent with the results from the paper "A Classification Engine for Image Ballistics of Social Data" by Giudice et al. \cite{giudice2017classification}.

\textbf{Filename.}
All files were renamed with  a specific naming convention regardless of the original image format and the platform used to upload the images. We then have a further investigation on how they name the uploaded images.

\textbf{Facebook naming convention:}
Facebook stores their photos using Haystack Photo Infrastructure \cite{beaver2010finding}, and when a user uploads a photo, it is assigned a unique 64-bit id. This id is then linked with the photo album to generate a link for the user to download the image. Before July 2012, Facebook named the image in a five-number pattern: $$aa\_bb\_cc\_dd\_ee\_n$$
in which $aa$ is the photo id (fbid), $bb$ is the album id, and cc is the profile id of the user who uploaded the picture; dd and ee were undisclosed. 
In July 2012, Facebook changed to a three-number pattern: $aa\_bb\_ee\_n$, i.e., the profile id was removed. 

\begin{figure}[t!]
    \centering
    \includegraphics[width=\linewidth]{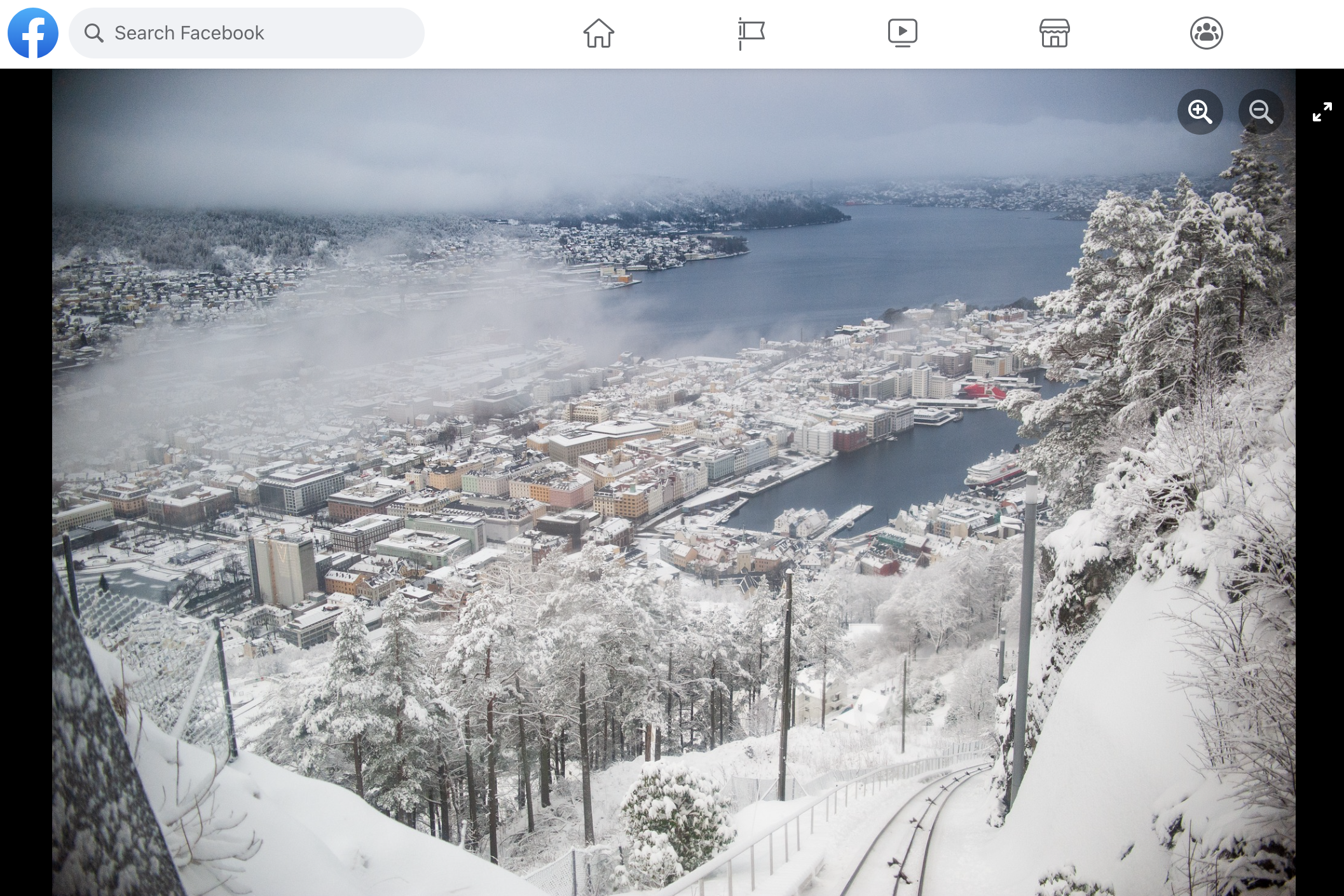}
    \vspace{-0.5cm}
    \caption{This photo on Facebook has fbid of 130119042955747. Users can access it via https://www.facebook.com/photo/?fbid=130119042955747. The downloaded photo is named: "280372071\_130119036289081\_1523944611184590851\_n.jpeg", which does not contain the fbid. Our findings point out that we can recover fbid just by having the 2nd part of the file name (130119036289081).}
    \label{fig:example2}
\end{figure}

At the time of this study, the name of uploaded picture on Facebook follows the four-number pattern: $$xx\_yy\_zz\_n\_aa$$
where $aa$ is still the photo id (fbid), and it can be used to access the image directly if the image privacy is set to public or if an account has been provided access through Facebook friendship. Figure~\ref{fig:example2} shows an example of how to access to a picture via its fbid.

If the image was downloaded through right clicking in the browser and selecting "Save As", users will only get the filename, and the fbid is removed: $$xx\_yy\_zz\_n$$ 

We observed in the 2475 image pairs, the first six digits of the 2nd part ($yy$) are identical to that of the fbid ($aa$). 
In order to explore this finding, we conducted another experiment by analysing additional 1998 images from the same Facebook account (that is the total number of the images from that user), as well as 200 images from a different account. The same pattern was confirmed.  

We propose a hypothesis that $aa$ can be recover by using only $yy$. To further test this hypothesis, we tried to compare the two numbers with some operations. By subtracting $aa$ from $yy$ (in the image in Figure~\ref{fig:example2}), we got $6666666$. When doing the same subtraction on other images, they returned numbers such as 3333333, 6666666, 9999999, 13333332, and 19999998, indicating a pattern in the relationship between the $yy$ and $aa$ where each difference between them was always a multiple of 3333333.

To see how many iterations it would take to enumerate the $yy$ until the same number as the $aa$ (each time by subtracting 3333333), we got the highest number was $213333312$ ($3333333 \times 64$). \textbf{This means one would only have to crawl through at most 64 iterations per image to find the real fbid from a filename.} 


\begin{table}[t!]
    \centering
    \caption{Examples of the relation between Facebook photo id (fbid - aa) and the 2nd part (yy) of the filename. We observed that all differences are multiples of 3333333.}
    \label{tab:example3}
    \resizebox{\columnwidth}{!}{%
        \begin{tabular}{lllc|lllc}
            \toprule
                yy & fbid (aa) & difference & factor & yy & fbid (aa) & difference & factor \\
            \midrule
                1507255906343130 & 1507256119676442 & 213333312 & 64 & 1507251046343616 & 1507251119676942 & 73333326 & 22 \\
                1507250843010303 & 1507250999676954 & 156666651 & 47 & 144877834756144  & 144877908089470  & 73333326 & 22 \\
                1507250899676964 & 1507251053010282 & 153333318 & 46 & 144877091422885  & 144877164756211  & 73333326 & 22 \\
                1507250853010302 & 1507250996343621 & 143333319 & 43 & 144877431422851  & 144877504756177  & 73333326 & 22 \\
                144877701422824  & 144877828089478  & 126666654 & 38 & 145475774696350  & 145475848029676  & 73333326 & 22 \\
                1507250789676975 & 1507250913010296 & 123333321 & 37 & 145475684696359  & 145475758029685  & 73333326 & 22 \\
                1507250933010294 & 1507251056343615 & 123333321 & 37 & 1507251473010240 & 1507251543010233 & 69999993 & 21 \\
                1507250846343636 & 1507250966343624 & 119999988 & 36 & 144877624756165  & 144877694756158  & 69999993 & 21 \\
                1507250849676969 & 1507250963010291 & 113333322 & 34 & 145475731363021  & 145475801363014  & 69999993 & 21 \\
                1507250949676959 & 1507251059676948 & 109999989 & 33 & 144877661422828  & 144877728089488  & 66666660 & 20 \\
                144877774756150  & 144877884756139  & 109999989 & 33 & 144877641422830  & 144877704756157  & 63333327 & 19 \\
                144877738089487  & 144877841422810  & 103333323 & 31 & 144877658089495  & 144877721422822  & 63333327 & 19 \\
                1507250816343639 & 1507250909676963 & 93333324  & 28 & 144877381422856  & 144877444756183  & 63333327 & 19 \\
            \bottomrule
        \end{tabular}%
    }
\end{table}

\subsection{Flickr}
\textbf{Exif.}
Flickr keeps all Exif data intact if the uploaded image is in JPEG. However, some entries are modified, deleted or updated if a RAW file is uploaded instead.

\textbf{IPTC Data.}
The JPEG images uploaded to Flickr show the same IPTC Digest hash as the original files. The RAW images uploaded to Flickr, however, have their Current IPTC Digest Hash modified, while still retaining the original IPTC Digest Hash.

\textbf{XMP Data.}
When uploading the JPEG files, Flickr retains all XMP data. Flickr retains some XMP data when uploading RAW images. The Original Document ID is retained and matches the original files.

\textbf{Compression}
Flickr re-compresses all the RAW files that were uploaded, but the JPEG files were not re-compressed.

\textbf{Resolution}
All images keep their original resolution after being uploaded to Flickr, regardless of the original file type.

\textbf{Filename and Naming Convention}
On Android and in the browser, while logged in to the account and downloading the images through the "Download Album" feature, Flickr keeps the original filenames as a prefix while adding an identifier following a specific pattern as a suffix:
$$filename\_identifier\_o.jpg$$
On iOS, the images are renamed with similar identifiers in a different order:
$$identifier\_filename\_o.jpg$$

When logged out of the account, i.e., not as the owner of the uploaded images, and download the images using the built-in Download feature, the images are named as $identifier\_filename\_o.jpg$, regardless the platforms  (Android, iOS or a web-browser).

Flickr allows users to choose a title when uploading an image. Changing this title or adding any title to the files uploaded from the iPhone application results in the title appearing as the prefix of the filename when the images are downloaded via the Download Album feature while logged in to the account. Also, while there is only one image in the album, if we use the Download Album feature, it will not include the title in the filename.

\begin{table}[h]
\caption{Example of how Flickr name image "r05d9d749t.jpg" after uploaded.}
\centering
\begin{tabular}{clll}
\toprule
Image Title     & Status & Platform & Downloaded File Name      \\ 
\midrule
No & Logged in & Android or PC & r05d9d749t\_52027420848\_o.jpg \\
No & Logged in & iOS & 52027420848\_r05d9d749t\_o.jpg \\
No & Logged out & Android, iOS, PC & 52027420848\_4efc66e8a4\_o.jpg \\
"Test" & Logged in & Android, iOS, PC & test\_52027420848\_o.jpg \\
"Test" & Logged out & Android, iOS, PC & 52027420848\_4efc66e8a4\_o.jpg \\
\bottomrule
\end{tabular}
\end{table}

Similar to the Facebook IDs, the Flickr IDs can be used to access publicly available images by modifying the ID parameter of the following URL: www.flickr.com/photo.gne?id=\textbf{52026001712}.
In addition to this, after accessing an image, the URL will change to display the ID of the user that uploaded the image: www.flickr.com/photos/\textbf{194832707}@\textbf{N04}/52026001712/.

\subsubsection{Decoding the IDs.}
In order to understand how FlickR generate the ID, we ran some additional experiments. 

The first learn if the uploaded locations could have any affect. To do that, images are re-uploaded while connected to a VPN (to Norway, Australia, US, and Vietnam). We have not observed any differences between the different locations. Interestingly, all the newly uploaded images have the ID starts with 5202, while the previous images starts with either 5197 or 5198, which seems to indicate that the ID number is increasing either based on time or the number of images on the platform. To test if time affects the ID numbers, we uploaded more images in the next day and observed the IDs are bigger than in the day before. Table~\ref{table: datetime} show some examples in our test.

\begin{table}[h]
\caption{Date and time of images uploaded to Flickr while waiting a certain amount of time between uploads}
\label{table: datetime}
\centering
\begin{tabular}{ccc|ccc}
\toprule
First day    & Date     & Time & Next day & Date & Time         \\ 
\midrule
52026301587                    & 04.24.22 & 09:08:51 p.m  & 52028946936 & 04.25.22 & 03:42:30 p.m. \\
52027901120                    & 04.24.22 & 09:08:53 p.m  & 52027942297 & 04.25.22 & 04:01:27 p.m. \\
52027390871                    & 04.24.22 & 09:08:54 p.m  & 52027951357 & 04.25.22 & 04:26:05 p.m. \\
52027420848                    & 04.24.22 & 09:08:52 p.m  & 52029084033 & 04.25.22 & 05:44:54 p.m. \\
52027606999                    & 04.24.22 & 09:08:51 p.m  & 52028186187 & 04.25.22 & 05:45:47 p.m. \\ 
\bottomrule
\end{tabular}
\end{table}

In addition to this test, older images on the platform are investigated. These images uploaded to the platform in 2017 have the same amount of digits in the ID, but they start with three instead of five: 36070463433, 36880336625, 36880336625. The ID of the image uploaded in 2012 starts with seven but has one less digit than the other IDs: 7542009332.

These results suggest that the IDs could be representing the total number of images on the platform. To further test this hypothesis, we ran the second additional experiment by starting at the ID of 52026001712, 3000 image URLs are crawled by enumerating the ID number and scraping the contents of the website to extract the date when the image was uploaded. Some images are deleted, which return a "Page not found" error, while some images are private. A total of 1035 images in the range from 52026001712 to 52026004712 were uploaded on April 24, 2022. One image with the ID 52026002941 was reportedly uploaded on April 23, another with ID 52026003572 was uploaded on April 25 and the other two were uploaded on February 26, 2022 (ID 52026001946 \& ID 52026001956). The "Date Taken" option on Flickr can be modified back to the year 1825, while "Date uploaded" can be changed to any date after the user has joined Flickr. Therefore, investigating time data of other images having adjacent IDs can give more confidence than just relying on these modifiable data alone.

\begin{table}[t!]
\caption{Findings Summary.}
\label{table:summary}
\centering
\begin{tabular}{c|p{3.8cm}|p{3.8cm}|p{2.4cm}}
\toprule
\textbf{Attribute} & \textbf{Flickr} & \textbf{Facebook} & \textbf{Google Photos} \\
\midrule
Exif & Only modified RAW images & Removed all fields except "Artist" and "Copyright"  tags	& Reserved \\
\midrule
IPTC & Reserved	& Removed all on the JPEG, retained some information on the RAW images	& Modified only if re-compression enabled\\
\midrule
XMP	& Only modified RAW images & Remove all & Modified only if re-compression enabled \\
\midrule
Resolution & Reserved	& Resized image larger than 2048px in either dimension	& Reserved \\
\midrule
Compression & Only compressed RAW images	& Recompressed regardless image format	&  Recompress if re-compression enabled \\
\midrule
File name & Renamed. Contains incremental number as id. This number can be used to retrieve the original uploaded time. & Renamed. The original photo ID (fbid) can be recovered from the file name. & Reserved \\
\bottomrule
\end{tabular}
\end{table}

\section{Potential Applications}
\label{sec:applications}
Image resolution could be a useful piece of data in digital fingerprinting if other social networks have similar criteria for resizing as the Facebook platform since they will resize all images that are above a certain size. This, combined with other data such as quantization tables could be useful regarding digital fingerprinting. The resolution would, however, not be effective in tracing images back to platforms like Google Photos or Flickr, since they do not alter the image dimensions in the upload process.

Filename modifications made by Flickr and Facebook, as long as they are not modified anywhere else, follow patterns that are specific to those platforms and can be of much help in determining the upload platform. On Flickr, the IDs present in the filenames can be used to access the image by using a specific URL that will take you to the corresponding image on the Flickr platform, but only if the image is publicly available. Whether the image is available or not, the ID number can be used to determine the date of upload by comparing it to other images in the same ID range. On Facebook, the same approach can be used as with Flickr by accessing the image (if it is publicly available) by using an ID that might be present in the filename. Whether or not this ID is present, another part of the filename can be used to reverse engineer the ID with little effort.

Considering the number of images that are uploaded to Flickr every day, estimating the date of upload for new images uploaded to Flickr is much easier than estimating images uploaded further back in time. Given the findings in this study, an accurate estimation of the date of upload of both current and historical Flickr images can be done. 

Since the image ID of Facebook images can be reverse engineered from the filenames of images downloaded from the Facebook platform, profile IDs and other obscured data might be possible to reverse engineer from the same filenames. If profile IDs can be reverse engineered from a filename, it might prove useful for tracing images back to the upload source in cases where the image is not accessible to the public. Obviously, we must also consider the privacy and ethical aspects.

\section{Conclusion}
\label{sec:conclusion}
The survey performed in this paper has shown that digital images undergo several changes in the upload process and that these changes vary greatly depending on the upload platform, as well as the file type, metadata, and other properties of the image file. Depending on the platform, metadata may be either unmodified, modified, or completely deleted. Depending on the file type of the original image and the settings on Google Photos, the file may or may not be re-compressed when it is uploaded to the social network. Facebook resizes all images above a set width or height of 2048 px, while the other social networks do not alter image dimensions up to at least 4928 px (the largest dimension used in this project) in width or height. Facebook and Flickr both alter the filenames in a way that is unique to each platform. These filenames contain data relevant to the image’s origin of upload, and in the case of Flickr, the upload date. These findings are summarised in Table~\ref{table:summary}. The proposed approaches based on these findings allow us to recover hidden information based barely on the file name of the uploaded photo.

\section*{Acknowledgement}
The results of this study are based on the consulting agreement between the Intelligent Information Systems (I2S) research group, University of Bergen, Norway and University of Science, VNU-HCM, Vietnam. 
The research was funded by European Horizon 2020 grant number 825469.

\bibliographystyle{splncs04}
\bibliography{bibliography}n

\end{document}